\documentclass[preprint]{aastex}
\usepackage{epsfig}

\slugcomment{Astrophysical Journal Letters (in press)}
\shorttitle{Energy release from evolving active regions}
\shortauthors{L. Vlahos, et al. }

\begin{document}
\title{Statistical properties of the energy release in emerging and evolving
active regions}

\author{Loukas Vlahos, Tassos Fragos, Heinz Isliker}
\affil{Department of Physics, University of Thessaloniki, 54006
Thessaloniki, Greece} \and
\author{Manolis Georgoulis}
\affil{Johns Hopkins University, Applied Physics Laboratory,
\\11100 Johns Hopkins Rd. Laurel, MD 20723-6099, USA}


\begin{abstract}
The formation and evolution of  active regions is an inherently
complex phenomenon. Magnetic fields generated at the base of the
convection zone follow a chaotic evolution  before reaching the
solar surface. In this article, we use a 2-D probabilistic
Cellular Automaton (CA) to model the statistical properties of the
magnetic patterns formed on the solar surface and to estimate the
magnetic energy released in the interaction of opposite
polarities. We assume that newly emerged magnetic flux tubes
stimulate the emergence of new magnetic flux in their
neighborhood. The flux-tubes move randomly on the surface of the
sun, and they cancel and release their magnetic energy when they
collide with magnetic flux of opposite polarity, or diffuse into
the ``empty'' photosphere. We assume that cancellation of magnetic
flux in collisions causes ``flares" and determine the released
energy as the difference in the square of the magnetic field flux
($E\sim B^2$). The statistics of the simulated ``flares" follow a
power-law distribution in energy, $f(E)\sim E^{-a},$ where
$a=2.2\pm 0.1$. The size distribution function of the simulated
active regions exhibits a power law behavior with index $k\approx
1.93 \pm 0.08$, and the fractal dimension of the magnetized areas
on the simulated solar surface is close to $D_F\sim 1.42\pm 0.12.$
Both quantities, $D_F$ and $k$, are inside the range of the
observed values.
\end{abstract}

\section{Introduction}
Many solar phenomena, such as coronal heating and solar flares,
are closely related to the evolution of  active regions. Active
regions are interpreted in this article as domains of strong
magnetic field on the solar surface.  The appearance of active
regions on the solar surface is the result of the complex
interplay between the buoyant forces and the turbulent convection
zone working in the solar interior \citep{Par79}. Convection zone
dynamics have been studied numerically \citep{Nor96}, but the
magneto-convection still remains in its infancy and many questions
remain unanswered (see review by \citet{Wei97}).

Many models have been suggested for the formation and evolution of
active regions, such as the rise of a kink-unstable magnetic flux
tube \citep{Mor92}, or the statistical description of the
dynamical evolution of large scale, two dimensional, fibril
magnetic fields \citep{Bog85}. \citet{Sch97} assumed that the thin
flux tubes that constitute the active region move and interact
only in the intergranular lanes, and form in this way a fractal
pattern. Several models have been developed using the anomalous
diffusion of magnetic flux in the solar photosphere in order to
explain the fractal geometry of the active regions \citep{Law91,
Sch92, Lsr93, Mze93}. Last, a percolation model was used to
simulate the formation and evolution of active regions
\citep{Wen92, Sei96}, which models the evolution of active regions
by reducing all the complicated solar MHD and turbulence to three
dimensionless parameters. This percolation model explains the
observed size distribution of active regions and their fractal
characteristics \citep{Meun99}.

Numerous observational studies have investigated the statistical
properties of active regions, using full-disc magnetograms and Ca
II plage regions observations from the Mount Wilson Observatory
and from the National Solar  Observatory (see review by
\citet{How96}). These studies have examined among other parameters
the size distribution of  active regions, and their fractal
dimension: {\em The size distribution function} of the newly
formed active regions exhibits a well defined power law with index
$\approx -1.94$, and active regions cover only a small fraction of
the solar surface (around $\sim 8\%$) \citep{Hzw93}.
 The \textit{fractal dimension} of the
active regions has been studied using high-resolution magnetograms
by \citet{Bak93}, and more recently by \citet{Meun99}. These
authors found, using not always the same method, a fractal
dimension $D_F$ in the range $1.3<D_F<1.8$.

In this article, we model the emergence and evolution of magnetic
flux on the solar surface  using a 2-D cellular automaton (CA),
following techniques developed initially by \citet{Sei96}. The
dynamics of this automaton is probabilistic and is based on the
competition between two ``fighting" tendencies:
\textbf{stimulated} or \textbf{spontaneous} emergence of new
magnetic flux, and the disappearance of flux due to
 \textbf{diffusion} (i.e.\ dilution below
observable limits),
 together with random \textbf{motion} of the
flux tubes on the solar surface. The basic new element we add to
the \citet{Sei96} model is that we keep track of the
\textbf{energy release} through flux cancellation (reconnection)
if flux tubes of opposite polarities collide. We concentrate our
analysis only on the newly formed active regions, since the old
active regions undergo more complicated behavior, the dipoles are
participating in large scale flows associated with differential
rotation and meridional flow \citep{Lei64, Wan94}.

\section{The  model}

We propose in this article that the main physical properties, as
derived from the observations of the evolving active regions, can
be summarized in simple CA rules.

A 2-D quadratic grid with $200 \times 1000$ cells (grid sites) is
constructed, in which each cell has four nearest neighbors. The
grid is assumed to represent a large fraction of the solar
surface. Initially, a small, randomly chosen percentage ($1\%$) of
the cells is magnetized (loaded with flux) in the form of
positively (+1) and negatively (-1) magnetized pairs (dipoles),
the rest of the grid points are set to zero. Positive and negative
cells evolve independently after their formation, but their
percentage remains statistically equal.

\noindent The dynamical evolution of the model is controlled by
the following probabilities:

\noindent  {\bf P}: The probability that a magnetized  cell is
stimulating the appearance of new flux at one of its nearest
neighbors. Each magnetized cell can stimulate its neighbors only
the first time step of its life. This procedure simulates the
stimulated emergence of flux which occurs due to the observed
tendency of magnetic flux to emerge in regions of the solar
surface in which magnetic flux had previously emerged.

\noindent {$\bf D_m$}: The flux of each magnetized cell has a
probability $D_m$ to move to a random neighboring cell, simulating
motions forced by the turbulent dynamics of the underlying
convection zone. If the moving flux meets oppositely polarized
flux in a neighboring cell, the fluxes cancel (through
reconnection), giving rise to a ``flare". If equal polarities meet
in a motion event, the fluxes simply add up.

\noindent  {\bf D}: The probability that a magnetized cell is
turned into non-magnetized in one time-step if it is next to a
non-magnetized cells.
This rule simulates the disappearing of flux below observational
limits due to dilution caused by diffusion into the empty
neighborhood.

\noindent {$\mathcal{E}$}: The probability that a non-magnetized
cell is turned into magnetized spontaneously, independently of its
neighbors, simulating the observed spontaneous emergence of new
flux.

\noindent Every newly appearing flux tube is accompanied by an
oppositely polarized mate, taking into account the fact that flux
appears always in the form of dipoles.

A detailed discussion of the connection between the parameters
$P,D_m,E$ and the physical mechanisms acting in the evolution of
active regions was established in the articles of \citet{Wen92}
and \citet{Sei96}.

\begin{figure}[ht]
\centering\epsfig{file=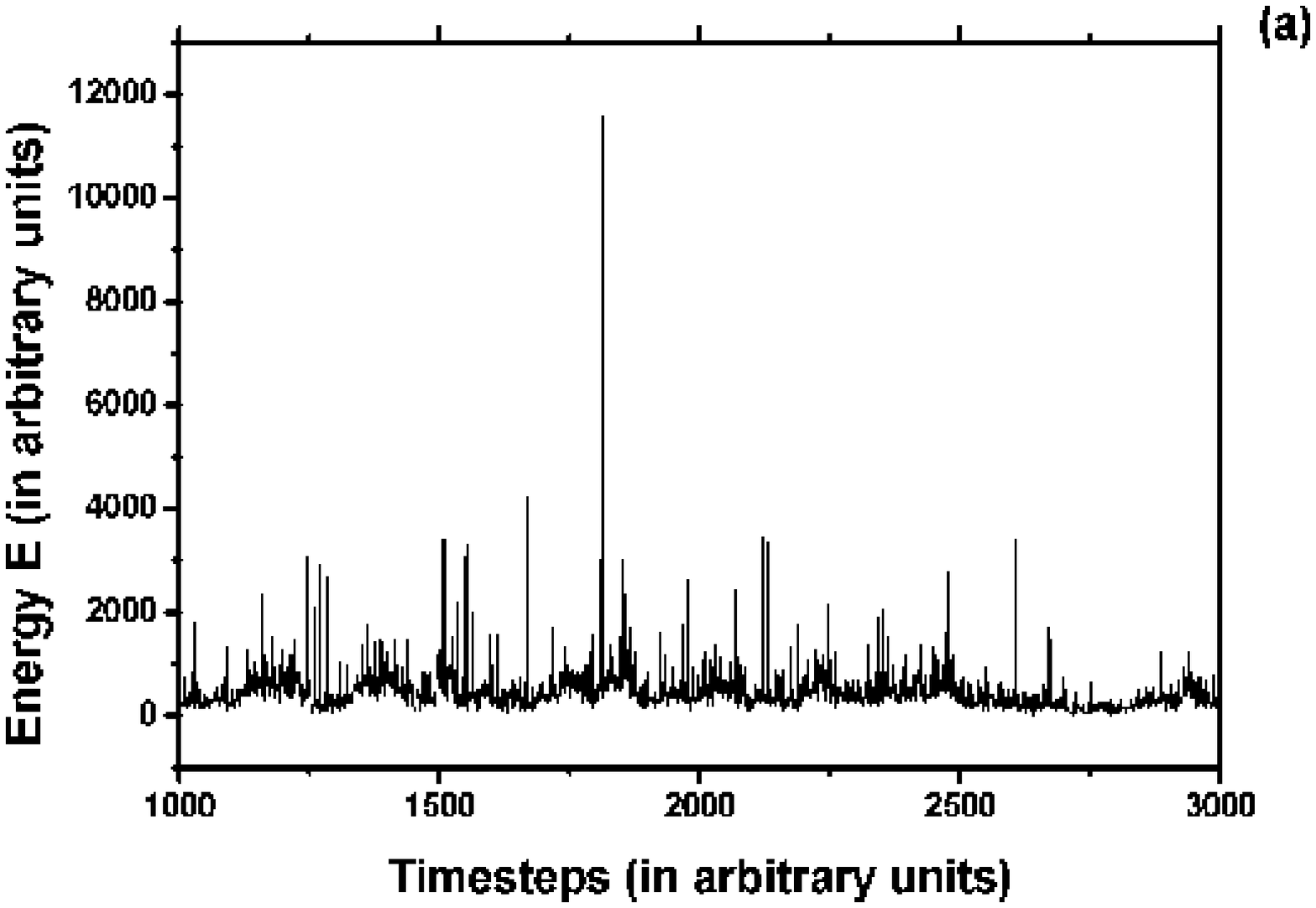,height=4.5cm}
\epsfig{file=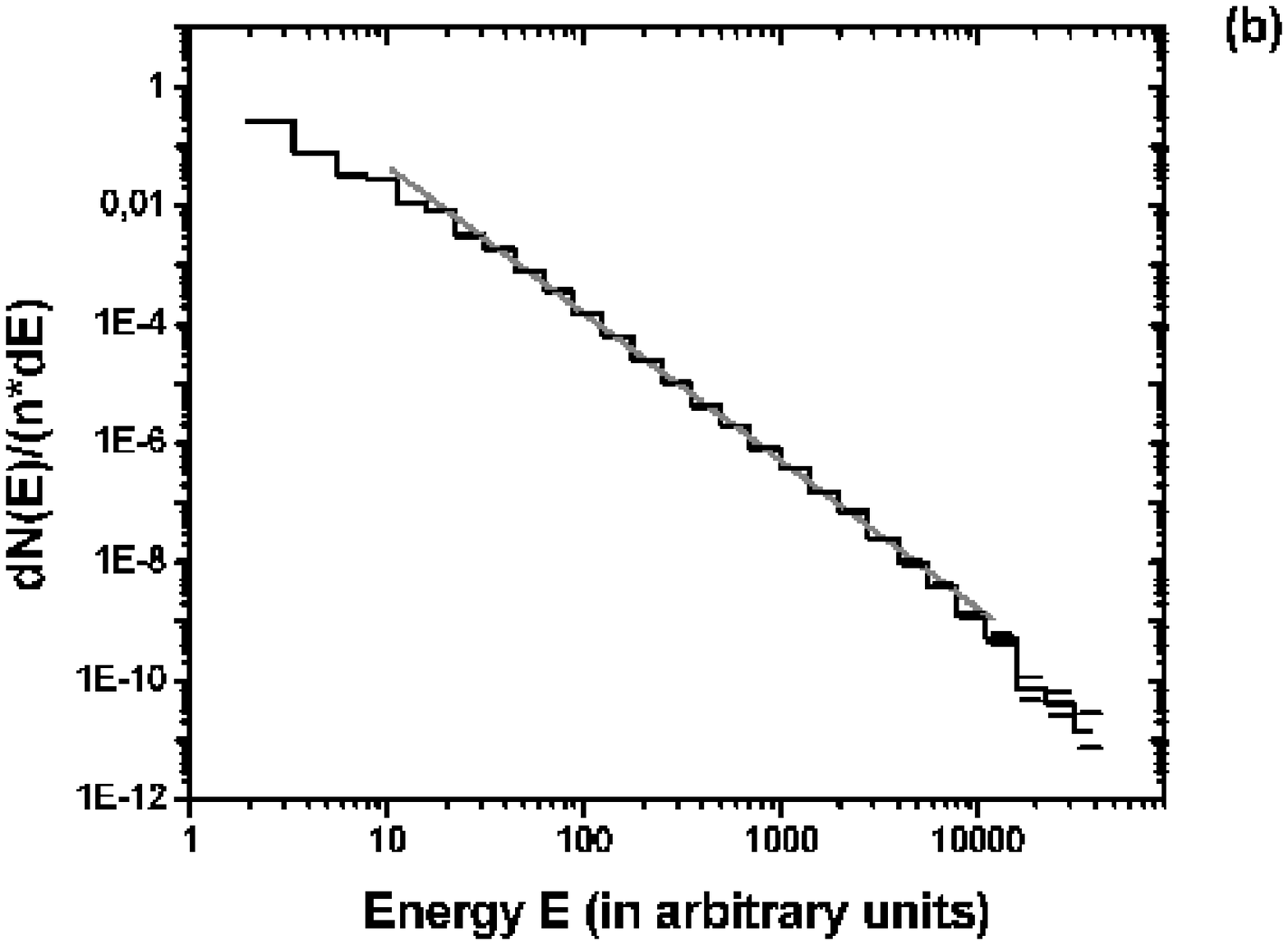,height=4.5cm} \caption{(a) The energy
released in the cancellation of magnetic flux as a function of
time, using the parameters $P=1.185, D=0.005, D_m=0.05,
 \mathcal{E}=10^{-6}$. (b) The energy distribution of the recorded
``flares". The power-law index is $2.2\pm 0.1$. \label{fig1}}
\end{figure}

\section{Results}
The parameters used for the results reported here are $P=0.185,
D=0.005,$ $D_m=0.05$ and $\mathcal{E}=10^{-6}$. They are chosen
such that, when following the evolution of our model and recording
the percentage of magnetized cells, we find that it takes around
1000 time steps before the percentage of active cells is
stabilized to a value which is close to the observed one (around
$8\%$). The size distribution of the simulated active regions can
be approximated by a power law fit of the form $N(s)\sim s^{-k} $,
with $k=1.93\pm 0.08$. Finally, we estimate the fractal dimension
$D_F$ of the set of magnetized cells with the box counting
algorithm (e.g. \citet{Fal90}), finding $D_F=1.42 \pm 0.12$.

\begin{figure}[ht]
\centering \epsfig{file=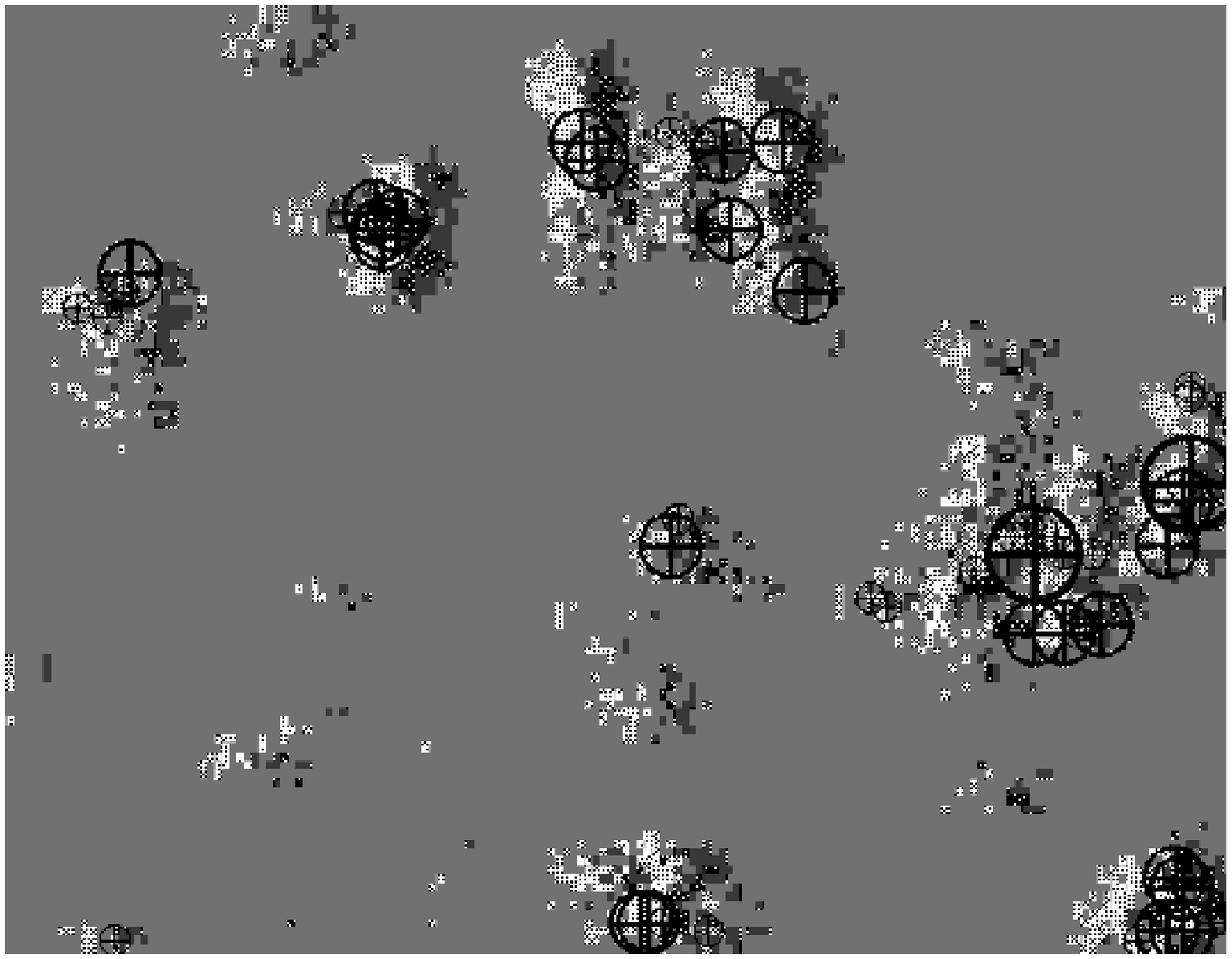,height=4.5cm}
\epsfig{file=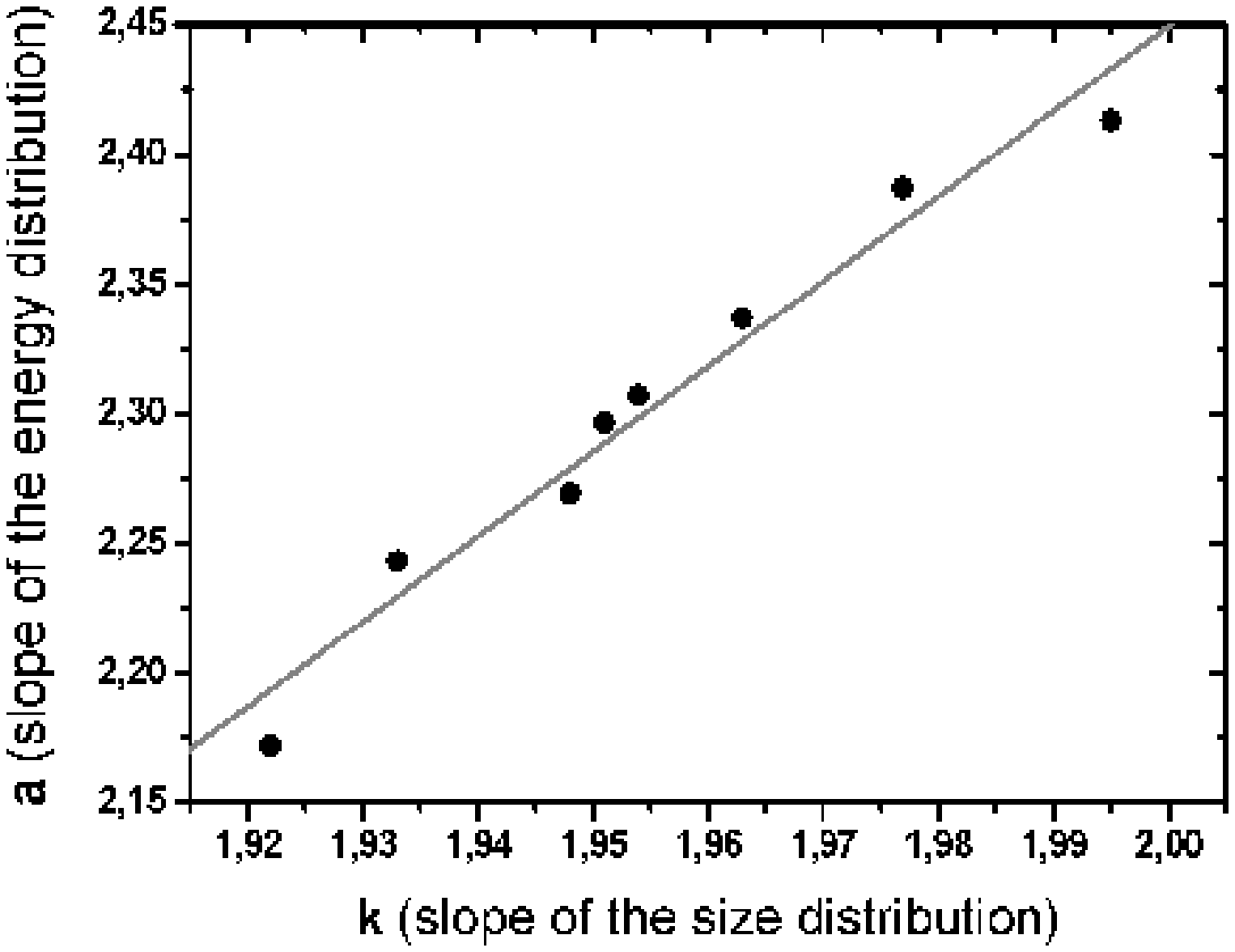,height=4.5cm} \caption{(a) A small portion of
the modeled grid is presented. The dark areas represent negative
and the white ones positive magnetic flux. The explosions
(``flares'') appear randomly at the interface of regions of
oppositely polarized magnetic flux. The circles represent  the
positions of the ``flares", with their radius being proportional
to the logarithm of the released energy. (b) There is a linear
dependence of the size distribution active region index k with the
energy release index a when the stimulation probability varies
$0.180<P<1.87, D=0.005, D_m=0.05$ and $\mathcal{E}=10^{-6}.$
\label{fig2}}
\end{figure}

The cancellation of magnetic flux due to collisions of oppositely
polarized magnetic flux tubes in motion events leads to the
release of energy, whose amount we assume to be proportional to
the difference in the square of the magnetic flux before and after
the event. In Fig.~  \ref{fig1}a, we plot the released energy
$E(t)$ as a function of time. Fig.~ \ref{fig1}b shows the energy
distribution of the recorded ``flares": It follows a power law,
$f(E)\sim E^{-a}$, with $a=2.2 \pm 0.1,$ for energies $E>20$. For
energies $E<20$ finite size effects must be expected to bias the
distribution, so we do not draw conclusions for the small
energies. It is important to note also that the power law in the
distribution of energy extends over three decades. In Fig.
\ref{fig2}a, we present a small portion of the grid. Dark areas
correspond to negative and bright ones to positive polarity. The
spatial locations of the ``flares" are marked with circles, with
the size of the circles proportional to the logarithm of the
locally released energy.

A variation of the parameters $P,D,D_m$ does not alter the power
law behaviours and the fractality, they seem to be
\textbf{generic} properties of the model. The exact values of the
power law indeces, $k,a,D_F$ depend though on the free parameters,
but remain inside the observed limits even for a large variation
of $P,D,D_m$. The results are also independent of $\mathcal{E}$ as
long as it remains small enough. In several independent
simulations, we used different values for the stimulation
probability $P,$ keeping $D$ and $D_m$ constant, and found that
$k$ and $a$ are closely correlated (see Fig.~ \ref{fig2}b).

\section{Summary and  Discussion}
We have introduced a probabilistic cellular automaton to explain
the formation and evolution of active regions and the associated
energy release. The rules used implement the competition between
the stimulated emergence of new magnetic flux and the gradual
diffusion, as well as the motion of flux across the photosphere.
We have compressed all the complicated physics involved in the
formation of active regions into three dimensionless parameters,
the probabilities $P$,  $D$, and $D_m$ of which actually only $P$
and $D_m$ play a decisive, major role on the results. The fourth
parameter $\mathcal{E}$ plays no significant role in the evolution
of the model (see Sec.\ 2).

Our main results are: (i) The model yields a power-law
distribution of the active region sizes, and the active regions
form a fractal set. (ii)  The collision and cancellation of
oppositely polarized magnetic   flux causes ``flares". The energy
distribution of the ``flares" follows a power-law distribution
with exponent $a=2.2 \pm 0.1$. Varying the parameters, the
power-law is conserved, just the power-law index change. (iii) The
``flares'' are distributed along the lines where regions of
opposite flux meet (see Fig.~ \ref{fig2}a). (iv) Using different
values for the stimulation probability $P$, we find a close
correlation between the size distribution exponent $k$ and the
energy release index $a$ (see Fig.~\ref{fig2}b).

The results derived from our model concerning size-distribution
and fractal dimension are in quantitative and qualitative
agreement with the existing observations and with the results of
the percolation models proposed previously \citep{Sei96,Meun99}.
The basically new result of our model is the connection of
photospheric dynamics to energy release. This primary energy
release through magnetic reconnection may well be the basic
physical cause for the various types of observed low atmosphere
emission events, such as coronal bright points
\citep{Par00,Kru98}, X-ray networks flares \citep{Krucker1997},
transition region impulsive EUV emission \citep{Benz2002}, and
H$\alpha$ flares (Ellerman bombs \citep{Georgoulis2002}). All
these emissions may play a role in coronal heating (this point is
still under strong debate), and in particular it was recently
claimed that impulsive, chromospheric EUV emissions are important
signatures of not just the coronal heating process but also of
mass supply to the corona \citep{Brown2000}.  Interesting to note
is that all the mentioned observed explosive phenomena exhibit
power-law distributions in energy, with slopes roughly in the
range between $1.5$ and $2.5$, coinciding with what we find here
for the primary photospheric energy release. We believe  that the
observed discrepancies in the indices (especially on $a$) are
directly related to the dynamical evolution of the active regions.
The predicted correlation between the power-law indices of the
distributions of the released energy and of the sizes of active
regions can be checked on existing observations. These results may
stimulate a new explanation for the statistical properties of
impulsive solar energy release events in the low atmosphere.

 We believe that CA models, when used in
conjunction with MHD numerical codes, can become a valuable tool
for the study of the statistical properties of active regions.

\begin{acknowledgements}
We thank our colleague Dr. A. Anastasiadis  for making several
suggestions and comments on the manuscript.
\end{acknowledgements}



\end{document}